\documentclass[%
preprint,
showkeys,
aps,
nofootinbib 
 ]{revtex4-1}
\usepackage{graphicx}
\usepackage{float} 
\usepackage{lipsum} 
\usepackage{hyperref}
\hypersetup{
  colorlinks=true,
  linkcolor=blue,
  filecolor=magenta,
  urlcolor=cyan,
  citecolor=cyan,
}
\usepackage{amsmath,amsfonts}
\usepackage{amssymb}
\PassOptionsToPackage{no-math}{fontspec} 
\usepackage{array}
\usepackage{braket} 
\usepackage{color} 
\usepackage{xcolor}
\usepackage{subfiles}

\usepackage[BoldFont,SlantFont,CJKnumber]{xeCJK}

\begin{document}

  \title{\huge Emergence of $\pi$ from Equatorial Quantum Localization\\
  \vspace{1cm} \color{teal}{\hrule height 4.0pt} }

\begin{center}
  \includegraphics[width=5.3cm]{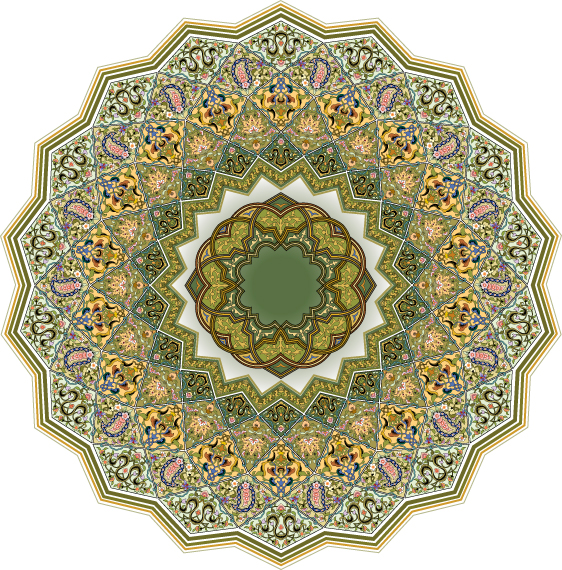}
\end{center}



\author{Bin Ye}
\author{Ruitao Chen}
\author{Lei Yin}
\email[Corresponding Author: ]{lei@scnu.edu.cn}
\affiliation{School of Materials and New Energy, South China Normal University, Shanwei 516625 China}

\begin{abstract}
    We present a genuinely non-radial quantum-mechanical route by which $\pi$ emerges from equatorial localization on the sphere. For the highest-weight branch of spherical harmonics, this localization is captured by a natural geometric rigidity index, whose exact finite-quantum-number value is a Wallis partial product. The mechanism is realized in two settings: the standard rigid rotor and the surface sector of a thin spherical shell, where radial freezing reduces the dynamics to the same angular problem. In the large-quantum-number limit, the probability cloud collapses toward the equator, the rigidity index approaches its classical value, and the Wallis formula is recovered through the correspondence principle. The result shows that Wallis-type structures in quantum mechanics can arise as exact signatures of semiclassical localization encoded by a simple geometric observable.
\end{abstract}


\maketitle

\tableofcontents

\section{Introduction and Main Results}
\label{sec:introduction}

Feynman once recalled being fascinated by the appearance of $\pi$ in physical formulas, especially in situations where no circle was immediately visible, and wondered where the hidden circle behind the number might be~\cite{Feynman1969WhatIsScience}. This elementary curiosity provides a natural point of entry for the present work. In the quantum setting considered here, the relevant question becomes: why should the number $\pi$ emerge from the classicalization of a quantum state? This question is appealing because it connects two themes that are usually treated separately: the geometry of quantum motion and one of the most celebrated constants in mathematics. We show that, for a highest-weight quantum state on the sphere, equatorial localization is measured by a simple geometric scalar whose exact finite-quantum-number value carries a Wallis structure; the corresponding infinite-product limit is precisely the Wallis formula for $\pi$. In this sense, the appearance of $\pi$ is not an external mathematical embellishment, but the signature of a correspondence-principle mechanism.

Earlier quantum-mechanical appearances of the Wallis formula were obtained mainly through radial or variational constructions. Friedmann and Hagen extracted the Wallis product from a variational treatment of the hydrogen atom in the large-angular-momentum regime, and subsequent works clarified that the phenomenon is not restricted to one particular trial function or to one particular Hamiltonian~\cite{FriedmannHagen2015,ChashchinaSilagadze2017,CorteseGarcia2018}. These works brought deserved attention to the surprising interplay between $\pi$ and quantum mechanics, but the mechanisms involved remained largely tied to radial equations, variational estimates, or model-specific constructions.

In a recent work~\cite{ye2026quantum}, we identified a related geometric mechanism in solvable radial systems, where a reciprocal radial observable acquires an exact finite Wallis structure and the large-quantum-number limit corresponds to the formation of a thin shell or a narrow annulus around a classical orbit. That result raises a natural question: is the Wallis structure fundamentally tied to radial localization, or can it also emerge from a genuinely non-radial quantum mechanism governed by angular geometry?

The present work answers this question in the affirmative. On $S^2$, the spherical-harmonic branch with maximal angular-momentum projection, $m=\ell$, carries a natural equatorial localization mechanism. The key observable is the equatorial rigidity index
\begin{equation}
  \mathcal R_m:=\left\langle \frac{R}{\rho}\right\rangle^{-1},
  \qquad
  \rho=R\sin\theta,
  \label{eq:Rm_intro}
\end{equation}
where $R$ is the fixed spherical radius and $\rho$ is the cylindrical radius with respect to a chosen symmetry axis. This observable compares a quantum state with the classical equatorial configuration in a probabilistic rather than pointwise sense. For the highest-weight spherical family, $\mathcal R_m$ is exactly
\begin{equation}
  \mathcal R_m
  =
  \frac{2}{\pi}
  \prod_{k=1}^{m}
  \frac{(2k)^2}{(2k-1)(2k+1)}.
  \label{eq:Rm_product_intro}
\end{equation}
Thus the Wallis formula follows from the correspondence-principle limit $\mathcal R_m\to 1$ as $m\to\infty$.

The construction has two main points of significance. First, it identifies a Wallis mechanism that is universal at the level of spherical kinematics, rather than tied to a specific radial equation. Second, it gives the $\pi$--quantum-mechanics connection a direct physical meaning: the Wallis product appears as the finite-quantum-number signature of equatorial classicalization.

We show that this spherical mechanism is realized in two settings. The first is the standard rigid rotor, the canonical realization of quantum motion on a sphere. The second is the surface sector of a thin spherical shell, a geometry naturally motivated by cold-atom bubble traps and shell-condensate systems~\cite{ZobayGarraway2001,Lundblad2019,Sun2018,TononiSalasnich2019}, where strong radial confinement reduces the dynamics to the same angular problem. In both cases, the highest-weight branch produces the same marginal density and the same rigidity index. The Wallis structure therefore reflects a common highest-weight spherical kinematics rather than a model-specific coincidence.

The paper is organized as follows. Section~\ref{sec:univ-spher-fram} develops the universal spherical framework, derives the rigidity index, and obtains the exact finite Wallis product. Section~\ref{sec:rigid-rotor-thin} shows how the same structure is realized in the rigid rotor and in the thin-shell surface problem. Section~\ref{sec:corr-princ-equat} analyzes the large-$m$ regime and closes the paper by interpreting the Wallis formula as the correspondence-principle limit of equatorial localization.

\section{Universal spherical framework and exact rigidity formula}
\label{sec:univ-spher-fram}

The central structure of the present work does not arise from a model-specific radial Schr\"odinger problem. Rather, it is rooted in a geometric mechanism on the sphere. Consider a sphere of fixed radius $R$, and let $\theta\in[0,\pi]$ be the polar angle measured from a chosen symmetry axis. The corresponding cylindrical radius is
\begin{equation}
  \rho=R\sin\theta .
  \label{eq:rho_def_sec2}
\end{equation}

At the classical level, the equator $\theta=\pi/2$ is distinguished by the maximal value $\rho=R$. For a quantum state, however, the relevant comparison with this classical reference is not the pointwise statement $\rho\to R$, but the concentration of the probability distribution near the equator. We therefore seek a geometric observable that measures, at the level of the probability cloud, how close a state is to ideal equatorial localization.

\subsection{Highest-weight family and equatorial reference}

The spherical mechanism is carried by the highest-weight angular family, namely the spherical-harmonic branch with maximal magnetic quantum number, $m=\ell$. Its angular dependence is
\begin{equation}
  Y_{mm}(\theta,\phi)\propto e^{im\phi}\sin^m\theta .
  \label{eq:Ymm_profile}
\end{equation}
After integration over the azimuthal angle $\phi$, the normalized polar marginal takes the form
\begin{equation}
  P_m(\theta)=N_m\sin^{2m+1}\theta,
  \qquad
  m=0,1,2,\dots,
  \label{eq:Pm_raw}
\end{equation}
with
\begin{equation}
  \int_0^\pi P_m(\theta)\,\mathrm d\theta=1 .
  \label{eq:Pm_norm}
\end{equation}
Introducing the sine integrals
\begin{equation}
  I_n:=\int_0^\pi \sin^n\theta\,\mathrm d\theta,
  \qquad n>-1,
  \label{eq:In_def}
\end{equation}
we obtain
\begin{equation}
  N_m=\frac{1}{I_{2m+1}},
  \qquad
  P_m(\theta)=\frac{\sin^{2m+1}\theta}{I_{2m+1}} .
  \label{eq:Pm_final}
\end{equation}
The exponent $2m+1$ has a transparent origin: the wave amplitude contributes $\sin^m\theta$, the probability density contributes its square, and the spherical measure contributes one additional factor $\sin\theta$. Thus the highest-weight branch naturally generates a profile centered on the equator and increasingly concentrated there as $m$ grows. For any function $f(\theta)$ depending only on $\theta$, the corresponding quantum average is
\begin{equation}
  \langle f(\theta)\rangle_m
  =
  \frac{
    \displaystyle
    \int_0^\pi f(\theta)\sin^{2m+1}\theta\,\mathrm d\theta
  }{
    \displaystyle
    \int_0^\pi \sin^{2m+1}\theta\,\mathrm d\theta
  }.
  \label{eq:average_def}
\end{equation}

\subsection{Geometric rigidity index}

Since a quantum state does not define a sharp trajectory, the relevant object is not the local variable $\rho$ itself, but its statistical distribution under $P_m(\theta)$. A natural dimensionless comparison with the equatorial value is the expectation of $R/\rho$, or equivalently of $\csc\theta$:
\begin{equation}
  \frac{R}{\rho}=\frac{1}{\sin\theta}=\csc\theta .
  \label{eq:R_over_rho}
\end{equation}
We define the equatorial rigidity index by
\begin{equation}
  \mathcal R_m:=\left\langle \frac{R}{\rho}\right\rangle_m^{-1}
  =\langle \csc\theta\rangle_m^{-1} .
  \label{eq:Rm_def}
\end{equation}
Because $\sin\theta\leq 1$ on $[0,\pi]$, one has $\csc\theta\geq 1$, and therefore
\begin{equation}
  0<\mathcal R_m\leq 1 .
  \label{eq:Rm_bound}
\end{equation}
The limiting value $\mathcal R_m=1$ corresponds to ideal equatorial localization, while $1-\mathcal R_m$ measures the defect from this ideal limit.

Using Eq.~\eqref{eq:Pm_final}, we find
\begin{align}
  \langle \csc\theta\rangle_m
  &=
  \int_0^\pi \csc\theta\,P_m(\theta)\,\mathrm d\theta
  \nonumber\\
  &=
  \frac{1}{I_{2m+1}}
  \int_0^\pi \sin^{2m}\theta\,\mathrm d\theta
  =
  \frac{I_{2m}}{I_{2m+1}} .
  \label{eq:csc_expect}
\end{align}
Hence
\begin{equation}
  \mathcal R_m=\frac{I_{2m+1}}{I_{2m}} .
  \label{eq:Rm_ratio}
\end{equation}
This is the basic structural result: the geometric measure of equatorial concentration reduces exactly to a ratio of two neighboring sine integrals.

It is useful to compare $\mathcal R_m$ with the angular-momentum alignment ratio. For the same highest-weight branch,
\begin{equation}
  \frac{\langle L_z\rangle}{\sqrt{\langle L^2\rangle}}
  =
  \frac{m\hbar}{\sqrt{m(m+1)}\,\hbar}
  =
  \sqrt{\frac{m}{m+1}},
  \label{eq:alignment_ratio}
\end{equation}
which also tends to $1$ as $m\to\infty$. This ratio measures angular-momentum alignment, whereas $\mathcal R_m$ measures coordinate-space equatorial localization. Only the latter carries the finite Wallis product.

\subsection{Finite-product structure and interpretation}

The even and odd sine integrals are
\begin{equation}
  I_{2m}
  =
  \sqrt{\pi}\,
  \frac{\Gamma\!\left(m+\frac12\right)}{\Gamma(m+1)}
  =
  \pi\,\frac{(2m)!}{2^{2m}(m!)^2}
  =
  \pi\,\frac{(2m-1)!!}{(2m)!!},
  \label{eq:I2m}
\end{equation}
and
\begin{equation}
  I_{2m+1}
  =
  \sqrt{\pi}\,
  \frac{\Gamma(m+1)}{\Gamma\!\left(m+\frac32\right)}
  =
  2\,\frac{2^{2m}(m!)^2}{(2m+1)!}
  =
  2\,\frac{(2m)!!}{(2m+1)!!} .
  \label{eq:I2m1}
\end{equation}
Substituting Eqs.~\eqref{eq:I2m} and \eqref{eq:I2m1} into Eq.~\eqref{eq:Rm_ratio} gives
\begin{equation}
  \mathcal R_m
  =
  \frac{2}{\pi}\,
  \frac{(2m)!!(2m)!!}{(2m-1)!!(2m+1)!!} .
  \label{eq:Rm_df}
\end{equation}
Using
\begin{equation}
  (2m)!!=\prod_{k=1}^{m}(2k),
  \qquad
  (2m-1)!!=\prod_{k=1}^{m}(2k-1),
  \qquad
  (2m+1)!!=\prod_{k=0}^{m}(2k+1),
  \label{eq:df_product}
\end{equation}
we obtain
\begin{equation}
  \mathcal R_m
  =
  \frac{2}{\pi}
  \prod_{k=1}^{m}
  \frac{(2k)^2}{(2k-1)(2k+1)} .
  \label{eq:Rm_product}
\end{equation}
If
\begin{equation}
  W_m:=\prod_{k=1}^{m}\frac{(2k)^2}{(2k-1)(2k+1)}
  \label{eq:Wm_def}
\end{equation}
denotes the finite Wallis partial product, then
\begin{equation}
  \mathcal R_m=\frac{2}{\pi}W_m .
  \label{eq:Rm_Wm}
\end{equation}
For the asymptotic analysis below, it is convenient to rewrite Eq.~\eqref{eq:Rm_ratio} as
\begin{equation}
  \mathcal R_m
  =
  \frac{\Gamma(m+1)^2}
  {\Gamma\!\left(m+\frac12\right)\Gamma\!\left(m+\frac32\right)} .
  \label{eq:Rm_gamma}
\end{equation}

The Wallis product is therefore not imposed from the outside. It is generated by the exact expectation value of a geometric localization observable that measures equatorial concentration of a highest-weight spherical state. The large-$m$ limit must be understood probabilistically: what approaches the equatorial configuration is not a pointwise orbit, but the quantum probability measure. The rigidity index $\mathcal R_m$ is the scalar measure of this concentration, and the finite Wallis product is its exact quantum signature.

\section{Rigid-rotor and thin-shell realizations}
\label{sec:rigid-rotor-thin}

We now show that the spherical mechanism established in Section~\ref{sec:univ-spher-fram} is realized in two physically distinct settings. The first is the standard rigid rotor, an exactly constrained surface system. The second is the surface sector of a thin spherical shell, where strong radial confinement produces an effective angular problem on the same sphere. The point of this section is not to rederive the rigidity index $\mathcal R_m$, but to identify the common highest-weight spherical kinematics underlying both realizations.

\subsection{Rigid rotor}

Consider a particle of mass $M$ constrained to move on a sphere of fixed radius $R$. The Hamiltonian is
\begin{equation}
  \hat H_{\rm rot}=\frac{\hat L^2}{2I},
  \qquad
  I=MR^2,
  \label{eq:Hrot}
\end{equation}
where $I$ is the moment of inertia. The eigenvalue equation
\begin{equation}
  \hat L^2Y_{\ell m}(\theta,\phi)
  =
  \hbar^2\ell(\ell+1)Y_{\ell m}(\theta,\phi)
  \label{eq:L2Ylm}
\end{equation}
gives the spectrum
\begin{equation}
  E_\ell=\frac{\hbar^2\ell(\ell+1)}{2I},
  \qquad
  \ell=0,1,2,\dots,
  \qquad
  m=-\ell,-\ell+1,\dots,\ell .
  \label{eq:Erot}
\end{equation}

The branch relevant for the present mechanism is the highest-weight family
\begin{equation}
  m=\ell\geq 0 .
  \label{eq:highest_weight}
\end{equation}
For fixed $\ell$, the rotor energy is independent of $m$. Choosing the state $m=\ell$ therefore selects the maximal angular-momentum projection along the chosen symmetry axis. This is the quantum family most directly associated with equatorial circular motion about that axis.

For this branch, the spherical harmonic takes the explicit form
\begin{equation}
  Y_{mm}(\theta,\phi)
  =
  (-1)^m
  \left[
    \frac{2m+1}{4\pi}\,
    \frac{(2m)!}{2^{2m}(m!)^2}
  \right]^{1/2}
  e^{im\phi}\sin^m\theta .
  \label{eq:Ymm}
\end{equation}
Thus the rigid rotor realizes exactly the highest-weight profile introduced in Section~\ref{sec:univ-spher-fram}. After integration over the azimuthal angle, the polar marginal is
\begin{equation}
  P_m(\theta)=\frac{\sin^{2m+1}\theta}{I_{2m+1}},
  \label{eq:Pm_rot}
\end{equation}
and hence the geometric rigidity index is the same quantity obtained in the universal spherical construction,
\begin{equation}
  \mathcal R_m
  =
  \left\langle \frac{R}{\rho}\right\rangle^{-1}
  =
  \langle \csc\theta\rangle^{-1},
  \qquad
  \rho=R\sin\theta .
  \label{eq:Rm_rot}
\end{equation}

The highest-weight choice also has a direct angular-momentum interpretation. On the state $Y_{mm}$,
\begin{equation}
  \hat L_zY_{mm}=m\hbar\,Y_{mm},
  \qquad
  \hat L^2Y_{mm}=m(m+1)\hbar^2\,Y_{mm}.
  \label{eq:LzL2}
\end{equation}
Therefore
\begin{equation}
  \frac{\langle L_z\rangle}{\sqrt{\langle L^2\rangle}}
  =
  \frac{m\hbar}{\sqrt{m(m+1)}\,\hbar}
  =
  \sqrt{\frac{m}{m+1}}
  \longrightarrow 1
  \qquad
  (m\to\infty).
  \label{eq:Lalign}
\end{equation}
The highest-weight rotor branch is therefore aligned in angular-momentum space and, as shown by $\mathcal R_m$, localized near the equator in coordinate space in the large-$m$ limit.

\subsection{Thin spherical shell}

We next consider a particle of mass $M$ in a spherically symmetric trapping potential $V(r)$ with a sharp minimum near $r=R_0$, a geometry naturally motivated by cold-atom bubble traps and shell-condensate systems~\cite{ZobayGarraway2001,Lundblad2019,Sun2018,TononiSalasnich2019}. In the thin-shell regime, the radial motion is strongly localized around $R_0$, and a one-body wavefunction may be factorized as
\begin{equation}
  \Psi(r,\theta,\phi)\approx f_0(r)\psi(\theta,\phi),
  \label{eq:factor_shell}
\end{equation}
where $f_0(r)$ is the radial ground mode.

Starting from the three-dimensional Hamiltonian
\begin{equation}
  \hat H
  =
  -\frac{\hbar^2}{2M}\nabla^2+V(r),
  \label{eq:H3d}
\end{equation}
and using
\begin{equation}
  \nabla^2
  =
  \frac{1}{r^2}\frac{\partial}{\partial r}
  \left(
    r^2\frac{\partial}{\partial r}
  \right)
  +
  \frac{1}{r^2}\Delta_{S^2},
  \label{eq:lap_spherical}
\end{equation}
projection onto the frozen radial mode gives an effective angular Hamiltonian. With the radial expectation value defined by
\begin{equation}
  \langle r^{-2}\rangle_0
  :=
  \int_0^\infty r^2\,dr\, |f_0(r)|^2\,r^{-2},
  \label{eq:rminus2_expectation}
\end{equation}
one obtains
\begin{equation}
  \hat H_{\rm surf}
  =
  E_r^{(0)}
  -
  \frac{\hbar^2}{2M}\langle r^{-2}\rangle_0\Delta_{S^2},
  \label{eq:Hsurf_general}
\end{equation}
where $E_r^{(0)}$ is the radial ground-state energy. Equivalently, one may introduce an effective surface radius by
\begin{equation}
  R_{\rm eff}^{-2}:=\langle r^{-2}\rangle_0 .
  \label{eq:Reff}
\end{equation}
In the ideal thin-shell limit, $R_{\rm eff}\to R_0$, and Eq.~\eqref{eq:Hsurf_general} reduces to
\begin{equation}
  \hat H_{\rm surf}
  =
  E_r^{(0)}
  -
  \frac{\hbar^2}{2MR_0^2}\Delta_{S^2} .
  \label{eq:Hsurf}
\end{equation}
Since $\hat L^2=-\hbar^2\Delta_{S^2}$, this may be written as
\begin{equation}
  \hat H_{\rm surf}
  =
  E_r^{(0)}+
  \frac{\hat L^2}{2MR_0^2} .
  \label{eq:HsurfL}
\end{equation}
Thus, up to the additive radial zero-point energy and the replacement $R\to R_0$, the ideal thin-shell problem is governed by the same angular Hamiltonian as the rigid rotor.

The surface eigenfunctions are again spherical harmonics,
\begin{equation}
  \psi_{\ell m}(\theta,\phi)=Y_{\ell m}(\theta,\phi),
  \label{eq:Ylm_shell}
\end{equation}
with energies
\begin{equation}
  E_\ell^{\rm surf}
  =
  E_r^{(0)}+
  \frac{\hbar^2\ell(\ell+1)}{2MR_0^2}
  \label{eq:Esurf}
\end{equation}
in the ideal thin-shell limit. Choosing again the highest-weight branch $m=\ell$, one obtains
\begin{equation}
  \psi_m(\theta,\phi)=Y_{mm}(\theta,\phi)\propto e^{im\phi}\sin^m\theta,
  \label{eq:highest_shell}
\end{equation}
so that the angular probability profile and the associated rigidity index coincide with those of the rigid rotor, with $R$ replaced by $R_0$.

The physical role of the shell realization is different from that of the rotor. The rotor is an exactly constrained surface system, whereas the shell geometry emerges as an effective surface theory after radial freezing. Nevertheless, the angular sector is the same, and it is this common surface kinematics that carries the rigidity observable and the associated finite-product structure.

\subsection{Common spherical kinematics}

Having separated the kinematically constrained rotor from the dynamically generated
thin-shell limit, we now collect the common angular structure shared by both systems.
In either realization, the highest-weight branch $m=\ell$ gives
\begin{equation}
  Y_{mm}(\theta,\phi)\propto e^{im\phi}\sin^m\theta ,
\end{equation}
so that the normalized polar marginal takes the universal form
\begin{equation}
  P_m(\theta)=\frac{\sin^{2m+1}\theta}{I_{2m+1}} .
  \label{eq:Pm_common_s3}
\end{equation}

The same geometric rigidity observable then follows, with $R$ replaced by $R_0$ in the shell case. The finite Wallis structure identified in Section~\ref{sec:univ-spher-fram} is consequently a property of highest-weight spherical kinematics rather than of either Hamiltonian separately.

\section{Correspondence principle and equatorial classicalization}
\label{sec:corr-princ-equat}

The exact finite-$m$ structure derived above acquires its physical meaning in the large-quantum-number regime. In both the rigid rotor and the thin spherical shell, the highest-weight branch becomes increasingly concentrated near the equator, and the rigidity index $\mathcal R_m$ tends to its classical value. The limit $m\to\infty$ is therefore not merely a formal asymptotic operation, but the correspondence-principle limit in which the quantum probability measure concentrates on the classical equatorial configuration.

\subsection{Equatorial localization}

From Section~\ref{sec:univ-spher-fram}, the normalized polar marginal is
\begin{equation}
  P_m(\theta)=\frac{\sin^{2m+1}\theta}{I_{2m+1}} .
  \label{eq:Pm_s4}
\end{equation}
Since $\sin\theta$ is maximal at $\theta=\pi/2$, the density is centered on the equator. To analyze the large-$m$ profile, write
\begin{equation}
  \theta=\frac{\pi}{2}+x,
  \qquad
  |x|\ll 1 .
  \label{eq:theta_local}
\end{equation}
Then
\begin{equation}
  \sin\theta=\cos x=1-\frac{x^2}{2}+O(x^4),
  \label{eq:sin_local}
\end{equation}
and therefore
\begin{align}
  \log P_m\!\left(\frac{\pi}{2}+x\right)
  &=
  \text{const}+(2m+1)\log(\cos x)
  \nonumber\\
  &=
  \text{const}-\frac{2m+1}{2}x^2+O(mx^4).
  \label{eq:logP_local}
\end{align}
Hence, near the equator,
\begin{equation}
  P_m\!\left(\frac{\pi}{2}+x\right)
  \approx
  P_m\!\left(\frac{\pi}{2}\right)
  \exp\!\left[-\frac{2m+1}{2}x^2\right].
  \label{eq:Pm_gaussian}
\end{equation}
Thus the highest-weight state is asymptotically Gaussian around the equator, with characteristic width
\begin{equation}
  \Delta\theta\sim (2m+1)^{-1/2}\sim m^{-1/2} .
  \label{eq:width_s4}
\end{equation}

Let $R_\ast$ denote the fixed spherical radius, with $R_\ast=R$ for the rigid rotor and $R_\ast=R_0$ for the thin-shell realization. The cylindrical radius is
\begin{equation}
  \rho=R_\ast\sin\theta .
  \label{eq:rho_Rstar}
\end{equation}
The equatorial configuration corresponds to
\begin{equation}
  \theta=\frac{\pi}{2},
  \qquad
  \rho=R_\ast .
  \label{eq:equatorial_ref}
\end{equation}
Thus the relevant classicalization statement is not the literal pointwise relation $\rho\to R_\ast$, but the concentration of the probability measure near the equatorial configuration $\rho=R_\ast$.

A complementary coordinate-space diagnostic is the height above the equatorial plane,
\begin{equation}
  z=R_\ast\cos\theta .
  \label{eq:z_coordinate}
\end{equation}
By symmetry, $\langle z\rangle=0$. Using the normalized density $P_m(\theta)=\sin^{2m+1}\theta/I_{2m+1}$, one obtains
\begin{equation}
  \langle z^2\rangle
  =
  R_\ast^2\langle \cos^2\theta\rangle
  =
  \frac{R_\ast^2}{2m+3} .
  \label{eq:z2_result}
\end{equation}
Therefore
\begin{equation}
  \frac{\sqrt{\langle z^2\rangle}}{R_\ast}
  =
  \frac{1}{\sqrt{2m+3}}
  \sim m^{-1/2} .
  \label{eq:z_width}
\end{equation}
Thus the vertical thickness of the probability band scales as $R_\ast m^{-1/2}$, in agreement with the Gaussian angular width in Eq.~\eqref{eq:width_s4}.

Figure~\ref{fig:Pm_theta} illustrates this localization by plotting the exact polar density $P_m(\theta)$ for several values of $m$. The narrowing around $\theta=\pi/2$ gives a direct visualization of the correspondence-principle limit.

\begin{figure}[!htb]
  \centering
  \includegraphics[width=0.52\columnwidth]{pm_theta}
  \caption{Exact polar density $P_m(\theta)=\sin^{2m+1}\theta/I_{2m+1}$ for several values of $m$. The distribution narrows around the equator as $m$ increases, illustrating the limit $\mathcal R_m\to 1$.}
  \label{fig:Pm_theta}
\end{figure}

\subsection{Rigidity asymptotics and scaling}

The rigidity index has the exact Gamma-function representation
\begin{equation}
  \mathcal R_m
  =
  \frac{\Gamma(m+1)^2}
  {\Gamma\!\left(m+\frac12\right)\Gamma\!\left(m+\frac32\right)} .
  \label{eq:Rm_gamma_s4}
\end{equation}
For large $m$, write
\begin{equation}
  \mathcal R_m
  =
  \frac{\Gamma(m+1)}{\Gamma\!\left(m+\frac12\right)}
  \cdot
  \frac{\Gamma(m+1)}{\Gamma\!\left(m+\frac32\right)} .
  \label{eq:Rm_split}
\end{equation}
Using the standard asymptotic expansions~\cite{DLMFGamma,tricomi1951asymptotic}
\begin{equation}
  \frac{\Gamma(m+1)}{\Gamma\!\left(m+\frac12\right)}
  =
  m^{1/2}
  \left(
    1+\frac{1}{8m}+\frac{1}{128m^2}+O(m^{-3})
  \right),
  \label{eq:gamma1}
\end{equation}
and
\begin{equation}
  \frac{\Gamma(m+1)}{\Gamma\!\left(m+\frac32\right)}
  =
  m^{-1/2}
  \left(
    1-\frac{3}{8m}+\frac{25}{128m^2}+O(m^{-3})
  \right),
  \label{eq:gamma2}
\end{equation}
we obtain
\begin{align}
  \mathcal R_m
  &=
  \left(
    1+\frac{1}{8m}+\frac{1}{128m^2}
  \right)
  \left(
    1-\frac{3}{8m}+\frac{25}{128m^2}
  \right)
  +O(m^{-3})
  \nonumber\\
  &=
  1-\frac{1}{4m}+\frac{5}{32m^2}+O(m^{-3}) .
  \label{eq:Rm_asymp}
\end{align}
Therefore the rigidity defect behaves as
\begin{equation}
  1-\mathcal R_m
  =
  \frac{1}{4m}+O(m^{-2}) .
  \label{eq:defect}
\end{equation}

The origin of this scaling is geometric. Near the equator,
\begin{equation}
  \sin\theta=\cos x=1-\frac{x^2}{2}+O(x^4),
  \qquad
  \csc\theta=\sec x=1+\frac{x^2}{2}+O(x^4) .
  \label{eq:csc_expand}
\end{equation}
Since $R_\ast/\rho=\csc\theta$, the local deviation from the equatorial value is quadratic in the angular displacement:
\begin{equation}
  \frac{R_\ast}{\rho}-1
  =
  \csc\theta-1
  \sim x^2 .
  \label{eq:csc_minus1}
\end{equation}
Together with $x\sim\Delta\theta\sim m^{-1/2}$, this gives
\begin{equation}
  1-\mathcal R_m\sim (\Delta\theta)^2\sim m^{-1} .
  \label{eq:defect_scaling}
\end{equation}
Thus the angular width and the rigidity defect describe the same localization process at two different levels: the width scales as $\Delta\theta\sim m^{-1/2}$, while the scalar defect is quadratic in this width and therefore scales as $m^{-1}$.

\subsection{Correspondence-principle interpretation}

The large-$m$ limit has the same physical meaning in both realizations of Section~\ref{sec:rigid-rotor-thin}. For the rigid rotor, the highest-weight branch becomes concentrated on a narrow band around the classical equatorial orbit. For the thin spherical shell, the same branch becomes a narrow circulating band on the shell surface. In both cases, the classicalization is probabilistic: the angular distribution sharpens around $\theta=\pi/2$, while the geometric scalar tends to its ideal equatorial value,
\begin{equation}
  \mathcal R_m\longrightarrow 1,
  \qquad
  m\to\infty .
  \label{eq:Rm_to_1}
\end{equation}

For the rigid rotor, this limit is accompanied by angular-momentum alignment,
\begin{equation}
  \frac{\langle L_z\rangle}{\sqrt{\langle L^2\rangle}}
  =
  \sqrt{\frac{m}{m+1}}
  \longrightarrow 1 .
  \label{eq:alignment_s4}
\end{equation}
This is consistent with the coordinate-space picture, but it is $\mathcal R_m$ that directly measures equatorial localization and carries the exact finite-product structure.

The essential point is therefore not merely that a Wallis-type product appears. The product is the exact finite-$m$ value of a geometric localization index for a highest-weight spherical state. Its large-$m$ limit is the correspondence-principle limit in which the probability measure concentrates on the equator. In this sense, the Wallis formula is recovered from a genuinely non-radial quantum mechanism with a direct geometric meaning.

This final observation also clarifies the broader significance of the result. Classicalization on the sphere can be encoded by a simple geometric observable, and its exact finite-quantum-number value already contains the Wallis structure. The Wallis formula is thus obtained not by inserting a mathematical identity into quantum mechanics, but as the limiting expression of equatorial quantum localization.


\providecommand{\href}[2]{#2}\begingroup\raggedright\endgroup

\end{document}